\documentclass[sigconf]{acmart}
\usepackage{enumitem}

\usepackage{multirow}
\usepackage{cleveref}
\usepackage{xspace}
\usepackage{tablefootnote}
\usepackage[caption = false, labelformat=simple]{subfig}

\crefname{table}{Table }{Tables}
\crefname{figure}{Figure }{Figures}
\crefname{section}{Section }{Sections}

\renewcommand{\shortauthors}{Fan Hu et al.}

\newcommand{\modify}[1]{{#1}}

\newcommand{\model}{TOSS\xspace}

\newcommand{\ModelBC}{\model$_{{[}BM25{]}+CodeBERT}$}
\newcommand{\ModelGC}{\model$_{{[}GraphCodeBERT{]}+CodeBERT}$}
\newcommand{\ModelCbC}{\model$_{{[}CodeBERT-bi{]}+CodeBERT}$}
\newcommand{\ModelJC}{\model$_{{[}Jaccard{]}+CodeBERT}$}

\newcommand{\ModelBJC}{\model$_{{[}BM25+Jaccard{]}+CodeBERT}$}
\newcommand{\ModelCbGC}{\model$_{{[}CodeBERT-bi + GraphCodeBERT{]}+ CodeBERT}$}
\newcommand{\ModelGBC}{\model$_{{[}GraphCodeBERT+BM25{]} + CodeBERT}$\xspace}
\newcommand{\ModelAllC}{\model$_{{[}ALL~Stage1~Methods{]}+CodeBERT}$}

\def\ie{\textit{i.e.}~}

\AtBeginDocument{%
  \providecommand\BibTeX{{%
    \normalfont B\kern-0.5em{\scshape i\kern-0.25em b}\kern-0.8em\TeX}}}

\copyrightyear{2023}
\acmYear{2023}
\setcopyright{acmcopyright}\acmConference[WSDM '23]{Proceedings of the Sixteenth ACM International Conference on Web Search and Data Mining}{February 27-March 3, 2023}{Singapore, Singapore}
\acmBooktitle{Proceedings of the Sixteenth ACM International Conference on Web Search and Data Mining (WSDM '23), February 27-March 3, 2023, Singapore, Singapore}
\acmPrice{15.00}
\acmDOI{10.1145/3539597.3570383}
\acmISBN{978-1-4503-9407-9/23/02}

\usepackage{tcolorbox}

\usepackage{caption}
\usepackage{amssymb}

\begin{document}

\title{Revisiting Code Search in a Two-Stage Paradigm}

\author{Fan Hu}
\authornote{Work done during internship at Microsoft Research Asia.}
\affiliation{\institution{School of Information, Renmin University of China}
\country{China}
}

\author{Yanlin Wang}
\authornote{Corresponding author: Yanlin Wang (wangylin36@mail.sysu.edu.cn).}
\authornote{Work done during the author’s employment at Microsoft Research Asia}
\affiliation{\institution{School of Software Engineering, Sun Yat-sen University}
\country{China}
}

\author{Lun Du}
\affiliation{\institution{Microsoft Research}
\country{China}
}

\author{Xirong Li}
\affiliation{\institution{MoE Key Lab of DEKE, Renmin University of China}
\country{China}
}

\author{\begin{tabular}{c}Hongyu Zhang\end{tabular}}
\affiliation{\institution{\begin{tabular}{c}The University of Newcastle\end{tabular} }
\country{Australia}
}

\author{Shi Han}
\affiliation{\institution{Microsoft Research}
\country{China}
}

\author{Dongmei Zhang}
\affiliation{\institution{Microsoft Research}
\country{China}
}

\renewcommand{\shortauthors}{Fan Hu et al.}
\begin{abstract}
With a good code search engine, developers can reuse existing code snippets and accelerate software development process. Current code search methods can be divided into two categories: traditional information retrieval (IR) based and deep learning (DL) based approaches. DL-based approaches include the cross-encoder paradigm and the bi-encoder paradigm. However, both approaches have certain limitations. The inference of IR-based  and bi-encoder models are fast, however, they are not accurate enough; while cross-encoder models can achieve higher search accuracy but consume more time. In this work, we propose TOSS, a two-stage fusion code search framework that can combine the advantages of different code search methods. TOSS first uses IR-based and bi-encoder models to efficiently recall a small number of top-K code candidates, and then uses fine-grained cross-encoders for finer ranking. Furthermore, we conduct extensive experiments on different code candidate volumes and multiple programming languages to verify the effectiveness of TOSS. We also compare \model with six data fusion methods. Experimental results show that TOSS is not only efficient, but also achieves state-of-the-art accuracy with an overall mean reciprocal ranking (MRR) score of 0.763, compared to the best baseline result on the CodeSearchNet benchmark of 0.713. Our source code and experimental data are available at: \url{https://github.com/fly-dragon211/TOSS}.

\end{abstract}

\ccsdesc[300]{Information systems~Retrieval models and ranking}

\keywords{Code search, two stage, information retrieval}

\maketitle


\section{Introduction} 

Code search is an important task in software engineering as it helps software development and maintenance~\cite{SingerLVA97,NieJRSL16}. With a well-developed code search system, developers can search for code snippets with natural language and reuse previously written code to accelerate software development.

\begin{figure}
    \subfloat[Bi-encoder. \label{fig:bi_encoder}]{\includegraphics[width=0.48\columnwidth]{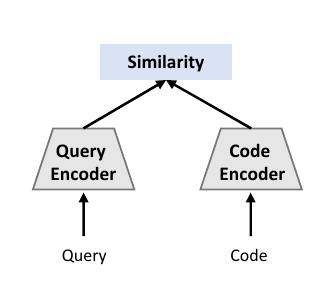}}
    \subfloat[Cross-encoder. \label{fig:cross_encoder}]{\includegraphics[width=0.48\columnwidth]{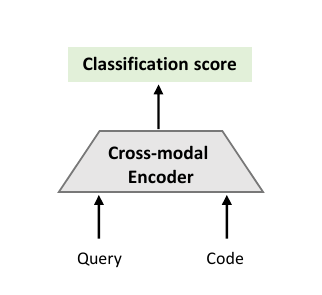}}  
    \caption{\textbf{The concept diagram of bi-encoder and cross-encoder code search models.} \modify{Bi-encoder  models are fast as the code embeddings can be pre-calculated offline. While cross-encoder models perform full-attention over the input pair of query and code, which could gain more information.}}
    \label{Figure:bi_vs_cross}
\end{figure}

Existing code search methods can be mainly divided into two categories, traditional information retrieval (IR) based (text matching) methods~\cite{nie2016query, yang2017iecs, rosario2000latent, hill2011improving, satter2016search, lv2015codehow, nie2016query,van2017combining} and deep learning (DL) based methods~\cite{GU2018Deep, Sachdev2018Retrieval,husain2019codesearchnet,20_code_bert,huang2021cosqa,iclr_graphcodebert,du2021single,shi2022enhancing,gu2022accelerating}. However, neither is ideal for applying the techniques to real world scenarios,  especially when the size of the searched codebase is huge. 
(1) IR-based methods are fast in terms of inference time, however, they are often not accurate. (2) On the contrary, DL-based methods can achieve higher performance, however, they are usually slower because of large models. Specifically, there are mainly two paradigms of model architecture in existing DL-based code search methods: cross-encoder paradigm \cite{20_code_bert,huang2021cosqa} and bi-encoder paradigm \cite{GU2018Deep, Sachdev2018Retrieval,husain2019codesearchnet,iclr_graphcodebert}. As described in \cref{Figure:bi_vs_cross}, cross-encoders perform full-attention over the input pairs of query and code, while bi-encoders map each input (query or code) independently into a dense vector space. The trade-off of efficiency and effectiveness also exist between these two paradigms: cross-encoders achieves better accuracy than bi-encoders. \modify{As the code embeddings can be pre-calculated and stored, bi-encoders are more efficient than cross-encoders.}
For many applications cross-encoders are not practical as they cannot produce separate embeddings for effective comparing or searching.

\vspace{5pt}

In this work, we systematically study the trade-offs between effectiveness and efficiency of existing code search methods. We aim to find out a combined solution that can not only keep the high search performance of cross-encoders but also reduce searching time. We select and reproduce nine code search methods with publicly available implementations on the CodeSearchNet~\cite{husain2019codesearchnet}  dataset. 
Firstly, we study the effect of different pre-processing operations in the code search task and find a proper pre-processing method that can boost the code search performance of text matching methods. 
Secondly, we systematically analyze the search performance, inference time, and recall results of these methods. We find that different methods are complementary to a certain extent. 
Therefore,  adopting the classic two-stage or multi-stage retrieval idea in the field of information retrieval, we design \textbf{\model}, a \textbf{T}w\textbf{O}-\textbf{S}tage fu\textbf{S}ion code Search framework. The first stage recalls a certain number (top-K) of code snippets using text matching or bi-encoder methods and the second stage uses fine-grained cross-encoders for reranking the code snippets recalled in the first stage.  
Finally, through extensive experiments, we find that improving recall diversity by combining the recalled code snippets from different first-stage models can improve the overall code search performance.  Our proposed method in the two-stage paradigm is not only efficient but also effective - it achieves the state-of-the-art results with an overall mean reciprocal ranking (MRR) score of 0.763, which is 7.1\% higher than the best baseline method. \modify{Besides, with the retrieval results of different models, one may consider the data fusion methods of IR, such as Max, Min, and so on. We also compare \model with the six data fusion methods described in \cref{related_fusion_method}.}
We position this work as a starting point research for exploring multi-stage retrieval in the code search task.

The contributions of this work can be summarized as:
\begin{itemize}
    \item We build a two-stage recall \& rerank framework for the code search task and adapt existing methods to this framework to improve  the effectiveness and efficiency of code search.
    \item We propose a multi-channel recall method that improves recall diversity and 
    code search performance in the two-stage paradigm. 
    \item Through extensive 
    experiments, we show the effectiveness of the proposed two-stage paradigm under different scenarios and data volumes.

\end{itemize}


\section{Related Work}\label{sec:rw}

\subsection{Code search methods}
Early explorations~\cite{nie2016query, yang2017iecs, rosario2000latent, hill2011improving, satter2016search, lv2015codehow,van2017combining} on code search mainly apply information retrieval (IR) techniques directly, which regard code search as a text matching task. Queries and code snippets are both regarded as plain text. The traditional text matching algorithms include BOW (bag-of-words)~\cite{schutze2008introduction}, Jaccard~\cite{jaccard1901etude}, TF-IDF (term frequency-inverse document frequency)~\cite{robertson1976relevance}, BM25 
~\cite{robertson2009probabilistic}, and extended boolean model~\cite{lv2015codehow}.

Since the cross-modal semantic gap 
is the major challenge for IR-based code search methods, researchers have explored many machine learning/deep learning based approaches~\cite{GU2018Deep, Sachdev2018Retrieval,husain2019codesearchnet,shuai2020improving,huang2020code,20_code_bert,huang2021cosqa,iclr_graphcodebert,sun2022importance} to capture the correlation between query and code from large-scale training data. 
Machine learning based code search models  \cite{GU2018Deep, Sachdev2018Retrieval} typically learn an embedding for query and code, and then calculate cross-modal similarity in a shared vector space. 
Early work on neural approaches to code search includes CODEnn \cite{GU2018Deep}, which uses RNN to jointly embed code snippets and natural language queries into a high-dimensional vector space. Husain et al. \cite{husain2019codesearchnet} proposed the CodeSearchNet benchmark and four baselines for code search, i.e., NBoW, 1D-CNN, biRNN, and SelfAtt. 
After the large-scale pre-training model BERT \cite{Devlin2019bert} was proposed, Feng et al. \cite{20_code_bert} proposed CodeBERT, which is  a  model pre-trained on unlabeled source code and comments.

For transformer-based code search methods, there are mainly two types of methods: bi-encoder and cross-encoder architecture. GraphCodeBERT \cite{iclr_graphcodebert} is a bi-encoder method, which encodes query and code into dense embeddings independently. 
CodeBERT \cite{20_code_bert} and CoCLR \cite{huang2021cosqa} are cross-encoders, where query and code are jointly encoded and we get a score that predicts whether a code answers a given query. Cross-encoder models process the query paired with each candidate code sequence, which is helpful to capture the relationship between two modalities (natural language and programming language). The concept diagram of bi-encoder and cross-encoder model is shown on \cref{Figure:bi_vs_cross}.

\subsection{Fusion methods} \label{related_fusion_method}
In the code search field, the fusion of IR, bi-encoder and cross-encoder code search methods are rarely explored. Some researchers have tried to use IR-based methods to improve the ML code search models. Sachdev et al. \cite{Sachdev2018Retrieval} used Word2vec method to get word embedding and aggregate representation of all the words with TF-IDF weights. Xie et al. \cite{xie2020source} calculated code similarity based on Siamese Neural Network. The weights of the word embeddings are fitted by TF-IDF. Recently, Gotmare et al. \cite{gotmare2021cascaded} proposed CasCode to fuse bi-encoder and cross-encoder models. However, CasCode only fuses the recall results of one bi-encoder model. In contrast, we propose a two-stage paradigm to fuse IR, bi-encoder and cross-encoder code search methods, which significantly improve the search performance and have a relatively fast search speed.

\modify{For fusion of the retrieval results of different source, one may consider the unsupervised data fusion method proposed in the information retrieval community, such as CombANZ \cite{fox1993combining, shaw1995combination}, Max, Min, CombMNZ \cite{fox1993combining, shaw1995combination}, CombSUM \cite{fox1993combining, shaw1995combination} and BordaCount \cite{aslam2001models}. The CombANZ method combines the similarity scores by computing the average of the non-zero scores. The Max and Min methods combine the similarity score sets by selecting the maximum or minimum one as the final score. The CombMNZ combines the similarity scores by multiplying the summation of all scores with the number of non-zero scores assigned to the method. The CombSUM combines the similarity scores by simply summing up their scores. The Borda count converts the similarity scores into ranks - codes with higher scores would obtain smaller ranks.  For each element, this method sums up the ranking points of an element given by a set of models. The ranking point of a code is defined as the substraction of the code’s rank in the list from the total number of codes in the codebase. }

\modify{Our two-stage code search paradigm \model is different from data fusion methods. Instead of fusing similarity score sets, \model first recalls  high-quality code snippets with first stage models, and then re-ranks this candidate set according to the second stage model. }

\begin{figure*}[t] 
\centering 
\includegraphics[width=0.98\textwidth]{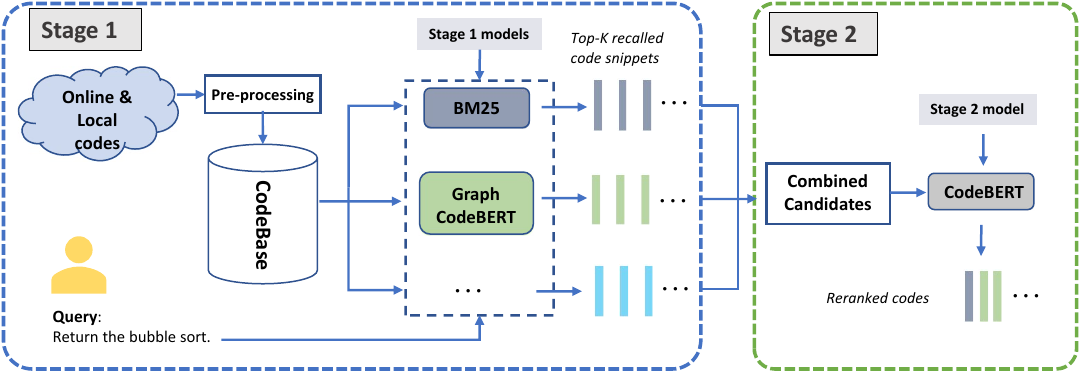}
\caption{The overall framework of our two-stage paradigm \model.}
\label{Figure:overall_framework}
\end{figure*}
\vspace{5pt}

\section{Framework}\label{sec:framework}

\subsection{Preliminaries}
Given a query $q$ and a codebase $C$, the goal of code search is to find the best code snippet that matches the query $q$ from the codebase $C$. Current code search methods can be uniformly formulated as follows:
\begin{equation}
\label{equ:overall_optimization}
    \max_{c \in C} \mathcal{M}(c, q),
\end{equation}
where $\mathcal{M}(\cdot, \cdot)$ is a similarity function to compute the match degree of the given query and a code snippet. The main difference of various code search methods is the design of the similarity function $\mathcal{M}(\cdot, \cdot)$.

For a traditional IR-based code search method, the similarity function $\mathcal{M}_{IR}(\cdot, \cdot)$ is designed as a text-based matching function. 
For a bi-encoder deep-learning model, we first transform the textual representations of the query $q$ and the code snippets $c$ to vector representations $e_q$ and $e_c$ by neural network encoders, and then calculate the similarity (or distance) measures in Euclidean space such as Cosine similarity or Euclidean distance to obtain the cross-modal similarity score $s_{bi}$. The calculation can be formalized as follows:
\begin{equation}
\begin{cases}
\mathbf{e_q} = \Gamma(q) \\
\mathbf{e_c} = \Gamma'(c), c \in C \\
s_{bi} = sim(\mathbf{e_q}, \mathbf{e_c})
\end{cases}
\end{equation}
where $\Gamma$ and $\Gamma'$ are two well-trained neural network encoders that can be the same networks or different networks according to the customized settings. 

For a cross-encoder deep-learning model, we concatenate the query and the code snippet, and input them to a neural network model to directly calculate the cross-modal similarity $s_o$:
\begin{equation}
    s_o = \mathcal{M}_{O} ({[q, <SEP>, c]}),
\end{equation}
where $<SEP>$ represents the separator between query and code. 
Since the embeddings of code snippets cannot be calculated offline, the search speed is always slower than the other two types of methods.

\subsection{Two-stage code search paradigm} \cref{Figure:overall_framework} depicts the overall framework of our two-stage code search paradigm \model.
Instead of directly selecting the best code snippet according to the overall target Eq.  \eqref{equ:overall_optimization}, the two-stage code search paradigm firstly recalls a candidate set of code snippets with a recall similarity function $\mathcal{M}_{recall}$ and then re-ranks the candidate set according to a ranking similarity function $\mathcal{M}_{rank}$. 
Formally:
\begin{align}
        &\rm{First\;Stage:} \quad C_{sub} = \mathop{\arg\max}_{C' \subset C, |C'| = K} \sum_{c \in C'} \mathcal{M}_{recall} (c, q),\\
        &\rm{Second\;Stage:} \quad c_* = \mathop{\arg\max}_{c \in C_{sub}} \mathcal{M}_{rank} (c, q),
\end{align}
where $C_{sub}$ is the candidate set that contains top $K$ code snippets with the recall similarity function $\mathcal{M}_{recall}$, and $c_*$ is the final selected code snippet after re-ranking. 
This code search paradigm enables a more accurate but time-consuming similarity function $\mathcal{M}_{rank}$,  especially when the code base is very large. 

In addition, by applying multiple search methods in the recall stage, we can incorporate the advantages of different methods and increase the diversity of the candidate set. \model uses $m$ fast first-stage models for recall, and then combine those returned code snippet sets as the final candidate set. It can be formalized as:
\begin{equation}
C_{sub}^{(i)} = \mathop{\arg\max}_{C' \subset C} \sum_{c \in C', |C'| = K} \mathcal{M}_{recall}^{(i)} (c, q), \; \rm{for} \; 0 < i \leq m
\end{equation}
\begin{equation}
    C_{combine} = \bigcup_i^m C^{(i)}_{sub},
\end{equation}
where $C_{sub}^{(i)}$ is the recalled top $K$ candidate set according to the $i$-th recall method, and $C_{combine}$ is the final candidate set. 

In conclusion, we expect this two-stage framework to accomplish two objectives: improved search performance and reduced search time. More specifically: (1) High-quality code snippets can be recalled in the first stage and ranked higher in the fine-grained second stage re-ranking. (2) The two-stage paradigm should considerably reduce research time by recalling a small set of code candidates for second stage re-ranking. 

\section{Experimental Design}\label{sec:exp_design}


\subsection{Datasets}


We conduct experiments on the widely used CodeSearchNet \cite{husain2019codesearchnet} dataset. 
Following Guo et al. \cite{iclr_graphcodebert}, we filter low-quality queries and expand the retrieval set to the whole code corpus. Note that we use the CodeSearchNet Python dataset, which contains 14,918 queries and 43,827 code candidates, in most of the 
experiments. 
We also generalize our findings to multiple programming languages with the full CodeSearchNet dataset in \cref{sec:compare_sota}.

\subsection{Baselines}
We select code search baselines to study based on their representativeness and their availability. The baselines that satisfy the following three criteria are chosen: 1) The source code is publicly available.  2) The overall model is adaptable to all the six programming languages in the CodeSearchNet dataset. 3) The paper is peer reviewed if it is proposed in a research paper. As a result, we select nine code search baselines and divide them into three categories: IR-based text matching methods (Jaccard~\cite{jaccard1901etude}, BOW~\cite{schutze2008introduction}, TFIDF~\cite{robertson1976relevance}, and BM25~\cite{robertson2009probabilistic}), cross-encoder deep learning based methods (CodeBERT~\cite{20_code_bert} and CoCLR~\cite{huang2021cosqa}), and bi-encoder deep learning based methods (CODEnn~\cite{GU2018Deep}, GraphCodeBERT~\cite{iclr_graphcodebert} and CodeBERT-bi).

\begin{itemize}
  \item \textbf{Jaccard.} We use \texttt{jaccard\_score}\footnote{\url{https://scikit-learn.org/stable/modules/generated/sklearn.metrics.jaccard_score.html}} from the \texttt{sklearn} python package to implement the Jaccard similarity method.

  \item \textbf{BOW.} We use \texttt{CountVectorizer}\footnote{\url{https://scikit-learn.org/stable/modules/generated/sklearn.feature_extraction.text.CountVectorizer.html}} to convert a query or code to a vector of token counts. The cross-modal similarity is then calculated by cosine similarity of code and query vectors.
  
  \item \textbf{TFIDF.} We use \texttt{TfidfVectorizer}\footnote{\url{https://scikit-learn.org/stable/modules/generated/sklearn.feature_extraction.text.TfidfVectorizer.html}} to convert a query or code to a vector of TF-IDF features. The cross-modal similarity  is calculated by cosine similarity of code and query TF-IDF features.
  
  \item \textbf{BM25.} We use the python package \texttt{Rank-BM25}\footnote{\url{https://pypi.org/project/rank-bm25/}} to implement the BM25 similarity.
  
  \item \textbf{CODEnn.} \modify{CODEnn is proposed in DeepCS \cite{GU2018Deep}, which jointly embeds code snippets and natural language descriptions into a high-dimensional vector space with LSTM. We use the implementation from their released repository}\footnote{\url{https://github.com/guxd/deep-code-search}}. 
  
  \item \textbf{CodeBERT.} A bi-encoder Transformer-based pre-trained model for programming language and natural language. We use the implementation from their released repository.\footnote{\url{https://github.com/microsoft/CodeBERT}}
  
  \item \textbf{CoCLR.} A CodeBERT-based contrastive learning method for code search. We use the implementation from their released repository.\footnote{\url{https://github.com/Jun-jie-Huang/CoCLR}}

 \item \textbf{GraphCodeBERT.} A dataflow aware pre-trained model based on CodeBERT. We use the implementation from their released repository.\footnote{\url{https://github.com/microsoft/CodeBERT/tree/master/GraphCodeBERT/codesearch}}
    
 \item \textbf{CodeBERT-bi.} CodeBERT bi-encoder variant described in the Appendix of the original paper~\cite{20_code_bert}, where CodeBERT first encodes query and code separately, and then calculates the similarity by dot product.

\end{itemize}

\subsection{Evaluation metrics} 
To evaluate the performance of code search models, we apply two popular 
evaluation metrics on the test set: MRR (Mean Reciprocal Ranking) and R@K (top-K accuracy, K=1, 5, 10, 100, 1000). They are commonly used for in previous code search studies~\cite{lv2015codehow, GU2018Deep, Sachdev2018Retrieval,husain2019codesearchnet,20_code_bert,huang2021cosqa,iclr_graphcodebert}. The higher the MRR and R@K values, the better the code search performance. In addition, we report per query search time as the efficiency measure.

\subsection{Experimental settings} 
For training, we use the default hyper-parameter settings provided by each method. For evaluation, the batch size is 256. For bi-encoder models, since source code can be processed offline, we do not include this computation cost in search time calculation. All experiments are conducted on a machine with Intel Xeon E5-2698v4 2.2Ghz 20-Core CPU and one Tesla V100 32GB GPU.




\section{Experimental Results}\label{sec:exp_results}

\begin{table}[t]
\centering 
\caption{\textbf{The code search performance (MRR) of different text matching methods with different text pre-processing tools.}}\label{tab:code_pre_processing}
\scalebox{0.8}{
\begin{tabular}{@{}cccccrrrrl@{}}
\toprule
\multicolumn{4}{c}{\textbf{Pre-processing tool}} & \multicolumn{1}{c}{\textbf{}} & \multicolumn{5}{c}{\textbf{Text matching method}} \\
\cline{1-4} \cline{6-10}
\textbf{SPS} & \textbf{DS} & \textbf{RS} & \textbf{POS} & \textbf{} & \textbf{Jaccard} & \textbf{BOW} & \textbf{TF-IDF} & \textbf{BM25} & \multicolumn{1}{l}{\textbf{Overall}} \\ \midrule
$\times$ & $\times$ & $\times$ & $\times$ &  & 0.1122 & 0.0949 & 0.1341 & 0.2914 & 0.1582 \\
\checkmark & $\times$ & $\times$ & $\times$ &  & 0.1823 & 0.1580 & 0.2274 & 0.4250 & 0.2482 $(56.9\% \uparrow)$ \\
$\times$ & \checkmark & $\times$ & $\times$ &  & 0.1353 & 0.1079 & 0.1251 & 0.3056 & 0.1685 $(6.5\% \uparrow)$ \\
$\times$ & $\times$ & \checkmark & $\times$ &  & 0.1842 & 0.1627 & 0.2326 & 0.4215 & 0.2503 $(58.2\% \uparrow)$ \\
$\times$ & $\times$ & $\times$ & \checkmark &  & 0.1162 & 0.1010 & 0.1384 & 0.2921 & 0.1619 $(2.39\% \uparrow)$ \\
\checkmark & \checkmark & $\times$ & $\times$ &  & 0.2321 & 0.1955 & 0.2194 & 0.4438 & 0.2727 $(72.4\% \uparrow)$ \\
\checkmark & \checkmark & \checkmark & $\times$ &  & 0.2366 & 0.2012 & 0.2257 & 0.4444 & 0.2770 $(75.1\% \uparrow)$ \\
\checkmark & \checkmark & \checkmark & \checkmark &  & \multicolumn{1}{r}{\textbf{0.2425}} & \multicolumn{1}{r}{\textbf{0.2220}} & \multicolumn{1}{r}{\textbf{0.2397}} & \multicolumn{1}{r}{\textbf{0.4523}} & \textbf{0.2891 $(82.8\% \uparrow)$} \\
\bottomrule
\end{tabular}
}
\end{table}
\vspace{5pt}

\begin{table*}[t]
\centering \setlength{\tabcolsep}{3.0pt}
\caption{\textbf{The code search performance of nine baseline methods. \modify{Per query time is the average retrieval seconds for randomly selected 100 queries. We repeat each time calculating experiment three times and report the mean and standard deviation. The cross-encoder DL models are time-consuming because the code features cannot be calculated offline. } } }
\label{tab:one_method_eval}
\begin{tabular}{@{}lrlrrrrrr@{}}
\toprule
\textbf{Model} & \multicolumn{1}{l}{\textbf{MRR}} & \textbf{Need training} & \multicolumn{1}{l}{\textbf{Per query time / s}} & \multicolumn{1}{l}{\textbf{R@1}} & \multicolumn{1}{l}{\textbf{R@5}} & \multicolumn{1}{l}{\textbf{R@10}} & \multicolumn{1}{l}{\textbf{R@100}} & \multicolumn{1}{l}{\textbf{R@1000}} \\ \midrule
\textit{\textbf{Text matching IR model}} & \multicolumn{1}{l}{\textit{\textbf{}}} & \textit{\textbf{}} & \textit{\textbf{}} & \multicolumn{1}{l}{} & \multicolumn{1}{l}{\textit{\textbf{}}} & \multicolumn{1}{l}{\textit{\textbf{}}} & \multicolumn{1}{l}{\textit{\textbf{}}} & \multicolumn{1}{l}{\textit{\textbf{}}} \\
Jaccard & 0.2425 & No & 0.0130 $\pm$ 0.0004 & 17.7 & 30.7 & 36.7 & 59.4 & 83.0 \\
BOW & 0.2220 & No & 0.0011 $\pm$ 0.0000 & 16.1 & 28.1 & 33.6 & 56.6 & 82.6 \\
TFIDF & 0.2397 & No & 0.0011 $\pm$ 0.0001 & 16.9 & 30.8 & 37.1 & 62.9 & 87.1 \\
BM25 & 0.4523 & No & 0.0062 $\pm$ 0.0003 & 35.6 & 56.4 & 63.4 & 81.0 & 92.0 \\ \midrule
\textit{\textbf{Bi-encoder DL model}} &  &  &  & \multicolumn{1}{l}{} & \multicolumn{1}{l}{} & \multicolumn{1}{l}{} & \multicolumn{1}{l}{} & \multicolumn{1}{l}{} \\
CODEnn & 0.1775  & Yes & 0.0033 $\pm$ 0.0001 & 11.1 & 23.9 & 30.7 & 57.3 & 82.3 \\
CodeBERT-bi & 0.6669 & Yes & 0.0021 $\pm$ 0.0003 & 57.4 & 77.9 & 83.3 & 94.6 & 98.8 \\
GraphCodeBERT & 0.6948 & Yes & 0.0048 $\pm$ 0.0002 & 59.3 & 82.1 & 87.3 & 96.5 & 99.1 \\ \midrule
\textit{\textbf{Cross-encoder DL model}} & \textit{\textbf{}} & \textit{\textbf{}} &  & \multicolumn{1}{l}{} & \multicolumn{1}{l}{\textit{\textbf{}}} & \multicolumn{1}{l}{\textit{\textbf{}}} & \multicolumn{1}{l}{\textit{\textbf{}}} & \multicolumn{1}{l}{\textit{\textbf{}}} \\
CodeBERT & \textbf{0.7015} & Yes & 802.43 $\pm$ 51.29 & 62.4 & 79.2 & 83.7 & 94.5 & 98.7 \\
CoCLR & 0.6349 & Yes & 766.27 $\pm$ 47.79 & 51.6 & 78.3 & 84.6 & 95.7 & 99.0 \\
\bottomrule
\end{tabular}
\end{table*}

\subsection{The impact of different pre-processing operations}
With the advent of deep learning approaches for code search, traditional IR-based code search approaches are often considered weak baselines \cite{husain2019codesearchnet, iclr_graphcodebert}. However, the IR-based approaches also have their own advantages, such as simplicity, training-free, and fast search speed, etc. Since text search methods are very sensitive to data pre-processing methods \cite{vijayarani2015preprocessing}, we investigate the effect of four data pre-processing operations that are commonly used in previous work on code search: 
\begin{sloppypar}
\begin{itemize}
  \item \textbf{SPS}: Split pascal and snake case. For example, it splits ``TwoStageMethod'' to ``Two Stage Method'' and ``vectorizer\_param'' to ``vectorizer param''. We use the regular expression (re) package\footnote{\url{https://docs.python.org/3/library/re.html}} to implement SPS. 
  
  \item \textbf{DS}: Delete the English stop-words, such as ``both'', ``more'', ``some'' and so on. We use stop-words implementation from the NLTK package.\footnote{\url{https://www.nltk.org/api/nltk.corpus.html}}  
  \item \textbf{RS}: Ronin Split, such as splitting "showtraceback" to "show trace back". We use ronin from spiral\footnote{\url{https://github.com/casics/spiral}}.

  \item \textbf{POS}: Part-of-speech restoration. For example, it restores ``configs'' to ``config''. We use the \texttt{WordNetLemmatizer} from the NLTK package\footnote{\url{https://www.nltk.org/api/nltk.stem.html}} for implementation.
\end{itemize}

\end{sloppypar}

To study the influence of different data pre-processing operations and find a suitable combination, we conduct a series of experiments on their combinations and evaluate the search performance (in MRR). As shown in  \cref{tab:code_pre_processing}, different data pre-processing methods can affect the overall performance by a large margin. The effect of using RS alone is the most obvious, with an average MRR improvement of 58.2\%. With all pre-processing methods, four text matching methods achieve best overall search performance.

In summary, we find that:
\begin{itemize}[leftmargin=15pt]
    \item Different text pre-processing methods can affect the overall performance by a noticeable margin. 
    \item SPS+DS+RS+POS is the recommended code pre-processing method, as the overall performance is best.  
\end{itemize}

\subsection{Analysis of baseline methods} \label{baseline_analyze}

In order to explore the complementary of various baselines, we study the nine baseline  methods in the following aspects:
\begin{itemize}
    \item \emph{Search performance measure}: MRR and R@k.
    \item \emph{Need training}: whether the method requires training data. 
    \item \emph{Search efficiency measure}: per query time.

\end{itemize}

As shown in \cref{tab:one_method_eval}, we observe that in general, most deep learning-based code search methods (CodeBERT, CodeBERT-bi, CoCLR, and GraphCodeBERT) perform better than IR-based text matching methods (Jaccard, BOW, TFIDF, and BM25). This result is consistent with previous work, as DL-based code search methods are data driven, which could learn embedding and find patterns from training data of different modalities (programming languages and natural languages). While text matching methods are rule based, which are sub-optimal compared with deep search methods especially in the code search scenario with different modalities. However, rule-based methods need no training data and less computing resource, which is more practical in industrial scenarios. The cross-encoder model CodeBERT performs the best among the nine methods in terms of search performance. However, in terms of time efficiency (inference time), bi-encoders and text matching methods greatly outperforms cross-encoder methods CodeBERT and CoCLR. Since cross-encoder methods cannot calculate code features offline, they require a lot of computing resources and may not be suitable for code search in industry scenarios with a large codebase.

\begin{figure*}[t]
\centering 
\includegraphics[width=0.8\textwidth]{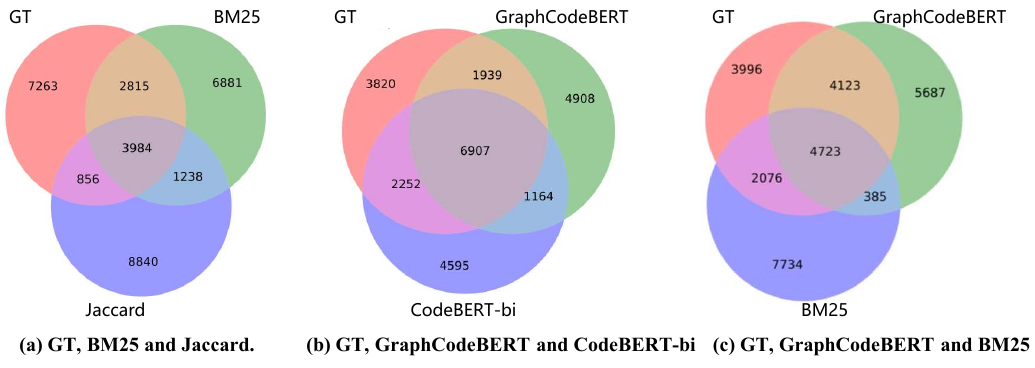}
\caption{\textbf{ Visualization results of the top-1 recalled samples based on the four baselines being used and ground truth (GT) in the CodeSearchNet python test set. } The diversity of recalled code candidates is higher for methods of different paradigms. The coincident number of recalls for fusing (GraphCodeBERT and BM25) is 5,108, which is less than text matching methods (BM25 and Jaccad) (5,222) and deep code search methods (GraphCodeBERT and CodeBERT-bi) (8,071). Besides, different methods can recall a part of unique ground truth code snippets. Best viewed in color.}
\label{Figure:stage1_venn}
\end{figure*}

With the above analysis, we conclude following key points:
\begin{itemize}[leftmargin=15pt]
    \item Most deep code search methods perform better than text matching methods, while the latter methods need no training data and less computing resource. 
    \item The cross-encoder model CodeBERT performs the best among the nine methods.
    \item Cross-encoder models are time consuming and not suitable for retrieval in large codebases.
    
    \item  Different baselines have their own characteristics. If we can integrate them well, the search performance and efficiency could be improved.
\end{itemize}

\begin{table*}[t]
\centering \setlength{\tabcolsep}{3.0pt}
\caption{\textbf{The two-stage code search performance of different baseline combinations.} \modify{Per query time is the average retrieval seconds for randomly selected 100 queries. We repeat each 
experiment three times and report the mean and standard deviation.} Stage 2 model: CodeBERT.}
\label{tab:model_combination_codebert}
\begin{tabular}{@{}llrrrrrr@{}}
\toprule
\multirow{2}{*}{\textbf{Combination}} & \multicolumn{3}{c}{\textbf{MRR}} & \multicolumn{1}{l}{} & \multicolumn{3}{c}{\textbf{Per Query Time / s}} \\ \cline{2-4} \cline{6-8} 
 & \multicolumn{1}{l}{\textbf{Top 5}} & \multicolumn{1}{l}{\textbf{Top 10}} & \multicolumn{1}{l}{\textbf{Top 100}} & \multicolumn{1}{l}{} & \multicolumn{1}{l}{\textbf{Top 5}} & \multicolumn{1}{l}{\textbf{Top 10}} & \multicolumn{1}{l}{\textbf{Top 100}} \\ \midrule
\textit{\textbf{Single stage1 model + CodeBERT}} & \multicolumn{1}{r}{} & \multicolumn{1}{r}{} & \multicolumn{1}{r}{} & \multicolumn{1}{r}{} & \multicolumn{1}{r}{} & \multicolumn{1}{r}{} & \multicolumn{1}{r}{} \\
\ModelBC & 0.5428 & 0.5852 & 0.6884 &  & 0.076 $\pm$ 0.002 & 0.127 $\pm$ 0.002 & 1.149 $\pm$ 0.006 \\
\ModelGC & 0.7392 & 0.7512 & 0.7589 &  & 0.539 $\pm$ 0.001 & 0.593 $\pm$ 0.001 & 1.619 $\pm$ 0.007 \\
\ModelCbC & 0.6911 & 0.7163 & 0.7566 &  & 0.163 $\pm$ 0.006 & 0.219 $\pm$ 0.005 & 1.210 $\pm$ 0.047 \\
\ModelJC & 0.3093 & 0.3530 & 0.5264 &  & 0.064 $\pm$ 0.001 & 0.123 $\pm$ 0.002 & 1.137 $\pm$ 0.009 \\ \midrule
\textit{\textbf{Multiple stage1 models + CodeBERT}} &  &  &  &  &  &  &  \\
\ModelBJC & 0.5706 & 0.6110 & 0.7058 &  & 0.133 $\pm$ 0.003 & 0.241 $\pm$ 0.001 & 2.278 $\pm$ 0.017 \\
\ModelCbGC & 0.7548 & 0.7588 & \textbf{0.7608} &  & 0.695 $\pm$ 0.004 & 0.806 $\pm$ 0.007 & 2.847 $\pm$ 0.010 \\
\textbf{\ModelGBC} & \textbf{0.7553} & \textbf{0.7595} & 0.7607 &  & 0.598 $\pm$ 0.001 & 0.712 $\pm$ 0.000 & 2.746 $\pm$ 0.009 \\
\ModelAllC & 0.7293 & 0.7462 & 0.7593 &  & 0.819 $\pm$ 0.003 & 1.043 $\pm$ 0.003 & 5.087 $\pm$ 0.049 \\
\bottomrule
\end{tabular}
\end{table*}
\vspace{1pt}

\begin{table}[tb!]
\centering \setlength{\tabcolsep}{7.0pt}
\caption{\modify{\textbf{MRR performance of different IR data fusion methods. Fusion data includes the similarity scores of GraphCodeBERT and BM25, and the re-ranking score of CodeBERT with top-K code snippets recalled.
} } }
\label{tab:compare_data_fusion}
\begin{tabular}{@{}lrrr@{}}
\toprule
\multirow{2}{*}{\textbf{Fusion methods}} & \multicolumn{3}{c}{\textbf{MRR}} \\ \cline{2-4} 
 & \textbf{Top 5} & \textbf{Top 10} & \textbf{Top 100} \\ \midrule
CombANZ & 0.5897 & 0.5846 & 0.4787 \\
Max & 0.4892 & 0.4923 & 0.4938 \\
Min & 0.5524 & 0.6286 & 0.6290 \\
CombMNZ & 0.5770 & 0.6154 & 0.5787 \\
CombSUM & 0.5825 & 0.6175 & 0.5787 \\
BordaCount & 0.5406 & 0.6061 & 0.6138 \\ \midrule
\textbf{TOSS} & \textbf{0.7553} & \textbf{0.7595} & \textbf{0.7607} \\ \bottomrule
\end{tabular}
\end{table}
\vspace{1pt}

\begin{figure*}
    \subfloat[MRR curves. \label{fig:mrr_vs_volume}]{\includegraphics[width=0.95\columnwidth]{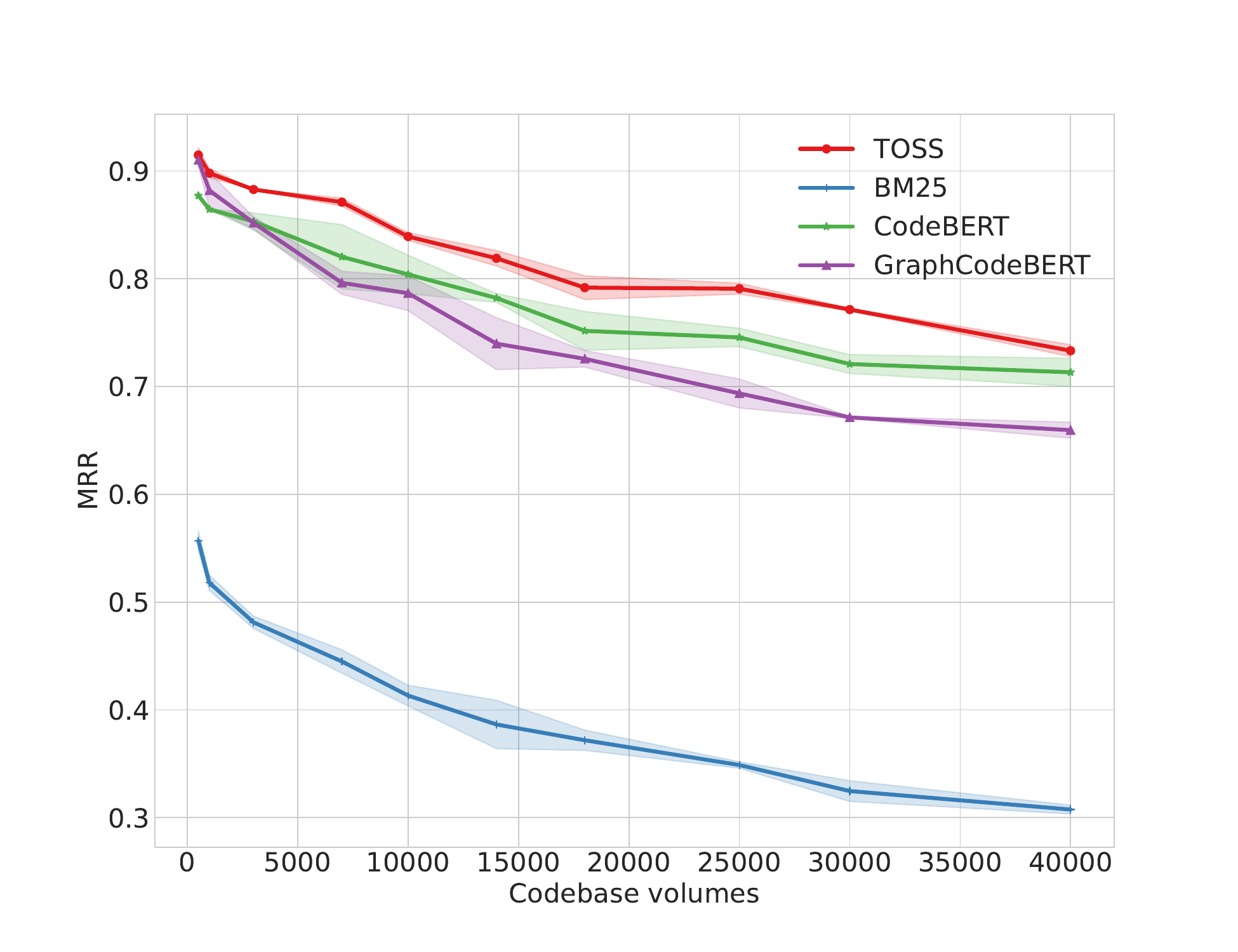}}
    \subfloat[Search time (log) curves. \label{fig:time_vs_volume}]{\includegraphics[width=0.97\columnwidth]{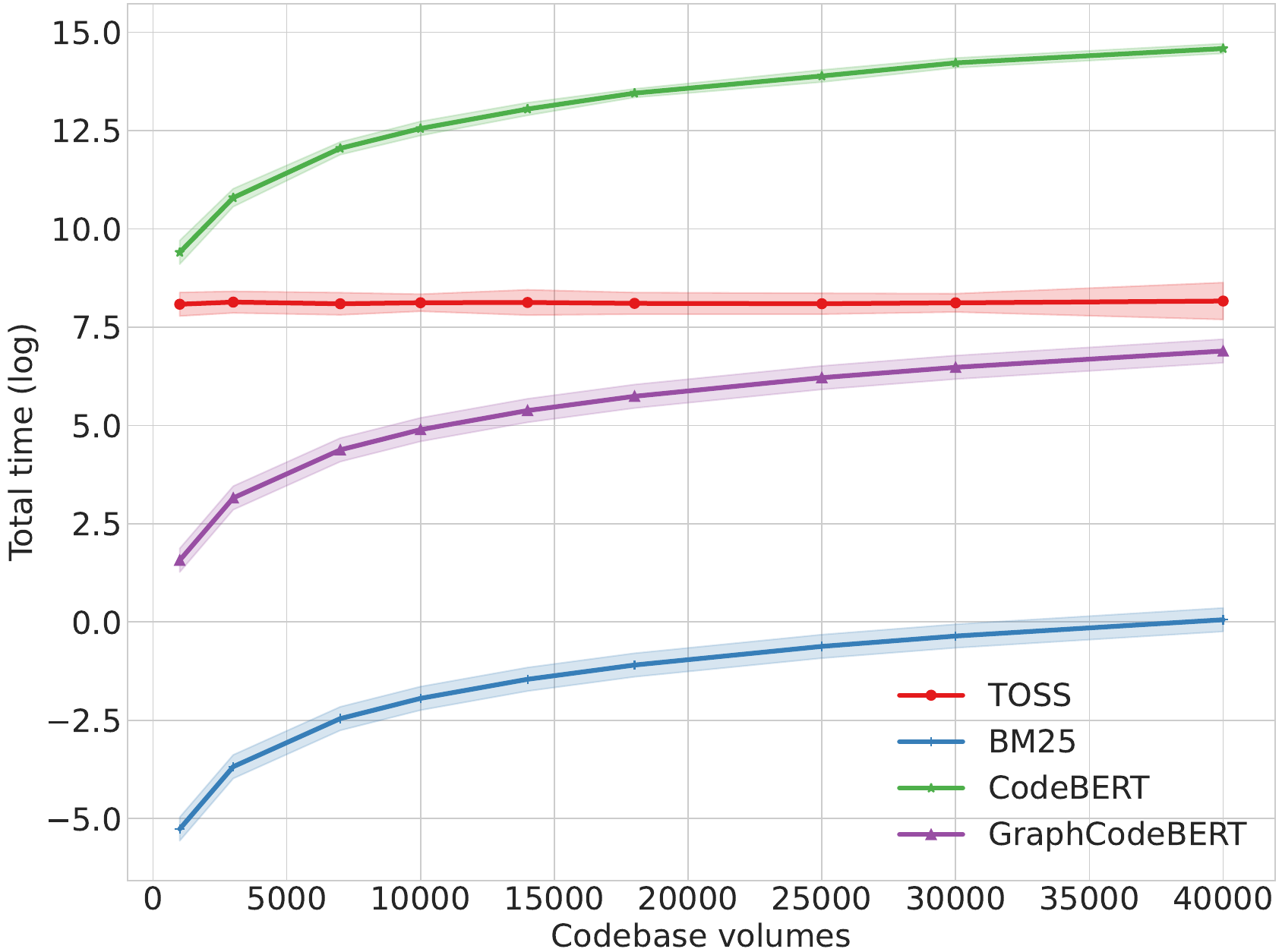}}  
    \caption{\textbf{Performance curves with different code volumes.} We set the code volume to be from 200 to 40000. \model refers to \ModelGBC. Since we randomly select a specific number of codes from the CSN python code candidates, we repeat each calculation three times and report the average results and the error bounds.
    }
    \label{Figure:mrr_time_curve}
\end{figure*}
\vspace{1pt}

\subsection{Effectiveness of the proposed paradigm \model } 

In this section, we want to explore the effectiveness of our proposed two-stage framework. The first-stage models are chosen from text matching and bi-encoder models.
The second-stage models are chosen from cross-encoder models\modify{, which have the slow time performance but full attention to the input pair of query and code}. We first explore the performance of a single first-stage model + second-stage model. Then we try multiple first-stage models + second-stage model. We find the best two-stage model \ModelGBC. Finally, we analyze the recall results of different first-stage models and find that multi-channel recalls have more overlap with the ground truth (GT), which leads to the boosting search performance of our proposed \model.

For first-stage models selection, 
\modify{We select 4 models with better accuracy and time performance from IR and bi-encoder models, \ie{GraphCodeBERT, CodeBERT-bi, BM25 and Jaccard}}. We put the top-K (K=5, 10, 100) code snippets recalled from the first-stage model into the second-stage model for re-ranking. We report MRR for search performance, per query time for efficiency measurement.

\begin{sloppypar}
The results of CodeBERT as the second-stage model are shown in \cref{tab:model_combination_codebert}.
We observe that compared with the model used alone, the \model method can significantly improve the code search performance. {\ModelGBC} achieves best performance. Compared with the best single-model CodeBERT, the MRR is improved by 8.6\%, and the retrieval time is reduced to 1/1400 of the original method. 
\end{sloppypar}

\begin{figure}[t]
\centering 
\includegraphics[width=0.50\textwidth]{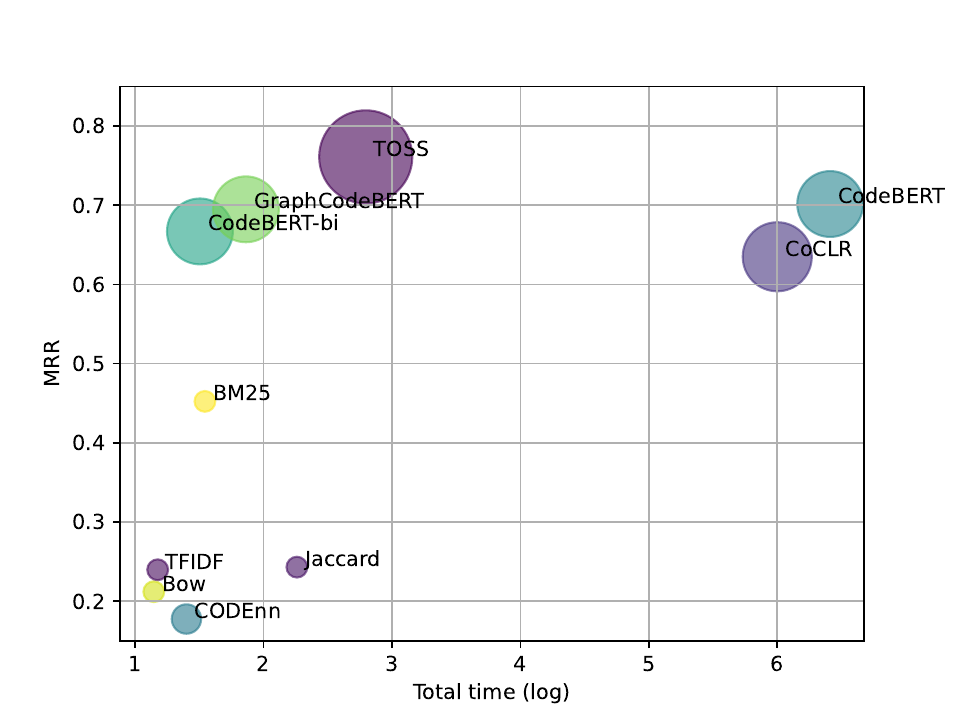} 
\caption{\textbf{ Visualization of the speed versus accuracy trade-off of nine baselines and our two-stage method.} Dataset: CodeSearchNet python test set. The area of the circle is proportional to the size of the model. The two-stage method \model refers to \ModelGBC.  With two-stage method, we are able to achieve top performance comparable to the best single model CodeBERT, while requiring substantially lesser inference time.}
\label{Figure:bubble} 
\end{figure}

\begin{sloppypar}
We also find that using multiple stage1 models for recall performs better than using single stage1 models. We attribute it to the complementarity of multiple models in the first stage, which can improve the recall ratio of the ground truth. To validate our assumption and explore how much complementarity the first-stage model recall have, we visualize the three combinations top 1 recall of the first-stage models, i.e., ground truth (GT) with two text matching methods (BM25 and Jaccad), two bi-encoder DL methods (GraphCodeBERT and CodeBERT-bi) and fusing methods (GraphCodeBERT and BM25). The result is shown in \cref{Figure:stage1_venn}. We obtain two keypoints. First, the recall results of different models vary greatly, and the intersection accounts for a small proportion of the total. Second, for the diversity of recalled results, the model from the two different paradigms are more diverse. The coincident number of recalls for fusing (GraphCodeBERT and BM25) is 5,108, which is less than text matching methods (BM25 and Jaccad) (5,222) and deep code search methods (GraphCodeBERT and CodeBERT-bi) (8,071). The performace of  {\ModelGBC} is better than that of 
{\ModelBJC} and {\ModelCbGC}, which proves that the fusion of two different channel recall is effective. The performance of {\ModelAllC} is not the best, which means that using too many first-stage models may introduce more negative samples, resulting in lower performance. The visualization of the speed versus accuracy trade-off of nine baselines and \model is shown on \cref{Figure:bubble}.
\end{sloppypar}

\begin{table*}[t]
\centering \setlength{\tabcolsep}{4.0pt}
\caption{\textbf{MRR performance on six languages of the CodeSearchNet dataset. TOSS refers to \ModelGBC.} }
\label{tab:compare_sota}
\scalebox{1.0}{
\begin{tabular}{@{}llllllll@{}}
\toprule
\textbf{Model / Method} & \multicolumn{1}{l}{\textbf{Ruby}} & \multicolumn{1}{l}{\textbf{JavaScript}} & \multicolumn{1}{l}{\textbf{Go}} & \multicolumn{1}{l}{\textbf{Python}} & \multicolumn{1}{l}{\textbf{Java}} & \multicolumn{1}{l}{\textbf{PHP}} & \multicolumn{1}{l}{\textbf{Overall}} \\ \midrule
BOW & 0.2303 & 0.1841 & 0.3502 & 0.2220 & 0.2447 & 0.1929 & 0.2374 \\
TF & 0.2390 & 0.2042 & 0.3625 & 0.2397 & 0.2620 & 0.2149 & 0.2537 \\
Jaccard & 0.2202 & 0.1913 & 0.3453 & 0.2425 & 0.2354 & 0.1822 & 0.2362 \\
BM25 & 0.5054 & 0.3932 & 0.5723 & 0.4523 & 0.4261 & 0.3352 & 0.4474 \\
CODEnn & 0.3420 & 0.3550 & 0.4951 & 0.1775 & 0.1083 & 0.1407 & 0.2698 \\
CodeBERT-bi & 0.6790 & 0.6200 & 0.8820 & 0.6669 & 0.6760 & 0.6280 & 0.6920 \\
GraphCodeBERT & 0.7030 & 0.6440 & 0.8970 & 0.6920 & 0.6910 & 0.6490 & 0.7127 \\ \midrule
\textbf{\model} & \textbf{0.7645 (8.7\%$\uparrow$)} & \textbf{0.6962 (8.1\%$\uparrow$)} & \textbf{0.9181 (2.4\%$\uparrow$)} & \textbf{0.7595 (9.8\%$\uparrow$)} & \textbf{0.7497 (8.5\%$\uparrow$)} & \textbf{0.6922 (6.7\%$\uparrow$)} & \textbf{0.7634 (7.1\%$\uparrow$)}
 \\
\bottomrule
\end{tabular}
}
\end{table*}

\modify{We also evaluate the search performance of \model with six IR data fusion methods presented in \cref{related_fusion_method}. 
We try to fuse the similarity scores of GraphCodeBERT and BM25 with the top-K (K=5, 10, 100) re-ranking CodeBERT similarity scores. Note that we use zero-one normalization to make the similarity scores comparable. The Mean Reciprocal Ranking (MRR) results are shown in \cref{tab:compare_data_fusion}. As we can see, \model outperforms all six fusion methods.
}


With above analysis, we conclude four key points:

\begin{itemize}[leftmargin=15pt]
    \item \model can significantly improve the code search accuracy with acceptable time consumption.
    \item The variant \ModelGBC~  \\ 
    achieves the best code search performance.
    \item Compared to single-channel method, multi-channel methods can recall more high-quality code candidates and boost the search performance.
    \item \modify{\model is more effective than the six selected data fusion methods.}
\end{itemize}

\subsection{Analysis of different code volumes}
In this section, to demonstrate the advantage of \model in different scenarios with different codebase sizes, especially when code search techniques are applied to real world applications with huge codebase, we study the impact of different code volumes for \model and baselines. 
We choose four methods for comparison, i.e., BM25 (text matching), GraphCodeBERT (bi-encoder), CodeBERT (cross-encoder) and our \ModelGBC (our two-stage paradigm). Each method is the best model in their respective categories. Due to the size limitation of the CodeSearchNet python dataset, the codebase cannot grow infinitely. We set the code volume to increase from 200 to 40,000. Since we randomly select a specific number of codes from the CodeSearchNet python code candidates, we repeat each calculation three times and report the average results and the error bounds.

We show the search performance MRR and per query time (log) with different code volumes in \cref{Figure:mrr_time_curve}. According to the results, we can observe that as the code volume increases, the search performance of each method decreases and the search time increases.

For MRR performance, as shown in \cref{fig:mrr_vs_volume}, among these models \model has consistently the best accuracy under different code volumes. Under 
small code volume, GraphCodeBERT performs well. While as the code volume increases, the search performance of GraphCodeBERT drops rapidly, which indicates that it is reasonable in our model \model to put GraphCodeBERT in the first stage and re-rank the small number of code candidates that are recalled.

For search time performance, as shown in \cref{fig:time_vs_volume}, text matching method BM25 has the shortest total time and cross-encoder method CodeBERT has the longest total time. Since the calculation time of the second stage is much larger than that of the first stage, and the calculation time of the second stage only depends on the size of k and has no dependency on the size of the original codebase, the search time of \model basically does not change much with the code volume. 

In summary, we obtain two key findings: 
\begin{itemize}[leftmargin=15pt]
    \item As the code volume increases, the search performance of each method decreases and the search time increases.
    \item The code search performance of \model is robust across different code volumes, and the computation time is fast and stable. Therefore, \model is a promising paradigm for real-world applications with large codebases.

\end{itemize}

\subsection{Compare with the baselines in multiple programming languages} \label{sec:compare_sota}

In this section, we compare the overall search performance of \model with baselines and study whether the performance improvement of \model still holds for other programming languages. \modify{As CodeBERT and CoCLR are time-consuming and doesn't suitable for large-scale retrieval, these two cross-encoder methods are not included.} We show the Mean Reciprocal Ranking (MRR) results in \cref{tab:compare_sota}. 
As we can see, \model outperforms all baseline methods on all the six programming languages. The average MRR of \model is 0.763, bring a 7.1\% gain to the best baseline method GraphCodeBERT. 

\section{Conclusion}\label{sec:conclusion}
In this paper, we present \model, a two-stage recall \& rerank framework for code search. It adapts existing methods to this framework to improve both  effectiveness and efficiency of code search. With multi-channel first stage method, we improve recall diversity and further improve code search performance in the two-stage paradigm. We evaluate different two stage model combinations and find the best two-stage model \ModelGBC, which means that GraphCodeBERT and BM25 are used as the first stage methods and CodeBERT is used as the second stage method.  We conduct extensive experiments on large-scale benchmark CodeSearchNet with six programming languages (Ruby, JavaScript, Go, Python, Java, PHP) and the results confirm its effectiveness in different scenarios and with different data volumes. 



\bibliographystyle{ACM-Reference-Format}
\balance
\bibliography{sample}

\end{document}